\documentclass[11pt]{article}
\usepackage{amssymb,amsmath,epsfig}

\begin{document}

\title{\bf Static Cylindrically Symmetric Interior Solutions in $f(R)$ Gravity}
\author{M. Sharif \thanks{msharif.math@pu.edu.pk} and Sadia Arif
\thanks{sadiaarif7@yahoo.com}\\
Department of Mathematics, University of the Punjab,\\
Quaid-e-Azam Campus, Lahore-54590, Pakistan.}

\date{}
\maketitle

\begin{abstract}
We investigate some exact static cylindrically symmetric solutions
for a perfect fluid in the metric $f(R)$ theory of gravity. For this
purpose, three different families of solutions are explored. We
evaluate energy density, pressure, Ricci scalar and functional form
of $f(R)$. It is interesting to mention here that two new exact
solutions are found from the last approach, one is in particular
form and the other is in the general form. The general form gives a
complete description of a cylindrical star in $f(R)$ gravity.
\end{abstract}

\section{Introduction}

Latest data from CMBR and supernovae surveys show that the energy
composition of the universe is as follows: $4\%$ ordinary brayonic
matter, $20\%$ dark matter and $76\%$ dark energy
\cite{8*}-\cite{11*}. Dark energy seems to be the most favorite way
to explain the accelerated expanding universe as indicated by recent
observations. However, the exact nature of dark energy is still
unknown. It is assumed to have a large negative pressure. There are
some problems in astrophysics and cosmology like the issues of dark
energy and dark matter where we experience severe theoretical
difficulties.

Many proposals are available in literature \cite{4}-\cite{7} to
generalize the theory of general relativity. The $f(R)$ theory of
gravity is the simplest generalization such that the Ricci scalar in
Einstein-Hilbert action is replaced by an arbitrary function of the
Ricci scalar. This generalization leads to complicated field
equations as well as of higher order. Due to higher order derivative
dependence, we expect to get more exact solutions than general
relativity. This theory explains the accelerated expansion of the
early universe as well as the current accelerated expansion of the
universe.

Recent literature indicates keen interest to investigate exact
solutions in $f(R)$ gravity. Multamaki and Vilja \cite{9} discussed
static spherically symmetric vacuum solutions. The same authors
\cite{10} extended it for perfect fluid and found that pressure and
density cannot be determined uniquely unless some conditions are
imposed. Caram\^{e}s and de Mello \cite{11} explored static
spherically symmetric vacuum solutions in higher dimensions.
Capozziello et al. \cite{12} found static spherically symmetric
vacuum solutions via N\"{o}ether symmetry approach. Hollenstein and
Lobo \cite{13} analyzed exact static spherically symmetric solutions
by coupling $f(R)$ gravity with nonlinear electrodynamics. Sharif
and Shamir \cite{18} explored static plane symmetric vacuum
solutions. The same authors \cite{19} found exact solutions for
Bianchi types I and V spacetimes in vacuum. Recently, Sharif and
Kausar \cite{20} discussed static spherically symmetric dust
solutions. The same authors \cite{21} also studied non-vacuum
solutions of Bianchi V$I_{0}$ universe. In a recent paper, Shojai
and Shojai \cite{22} found some static spherically symmetric
solutions for perfect fluid satisfying the physical acceptability
criteria.

The study of exact solutions in $f(R)$ gravity is mostly restricted
to spherical symmetry either vacuum or non-vacuum. One can compare
the results with solar system observations based on the
Schwarzschild solution. The physical acceptability of these
solutions can also be discussed. Azadi et al. \cite{31} investigated
static cylindrically symmetric vacuum solutions. Momeni and
Gholizade \cite{32} found static cylindrically symmetric vacuum
solutions with constant scalar curvature applicable to outside of a
string. It would be interesting to extend the study of exact
solutions for cylindrically symmetric non-vacuum solutions. Delgaty
and Lake \cite{33} gave the physical acceptability criteria for an
exact solution in GR as follows:
\begin{itemize}
\item  isotropy of pressure;
\item  regularity at the origin;
\item  positive definiteness of the energy density and pressure at the
origin;
\item  vanishing of pressure at some finite radius;
\item  the monotonic decrease of the energy density and pressure with increasing
radius;
\item  subluminal sound speed, i.e., (\textbf{${v_{s}}^{2}=\frac{dp}{d\rho}<1$}).
\end{itemize}
For $f(R)$ gravity, only the first and last two conditions are
accounted.

In this paper, we construct some static cylindrically symmetric
interior solutions in $f(R)$ gravity. We consider three types of
assumptions depending upon model parameters to explore these
solutions in tabular form and discuss them by using the physical
acceptability criteria. In particular, we are interested in
physically acceptable solutions of a cylindrical star. We discuss
the dependence of $f(R)$ functions on the scalar curvature for these
solutions. The outline of the paper is as follows: In section
\textbf{2}, we give a brief overview of the metric $f(R)$ gravity
and formulate the field equations of cylindrically symmetric
spacetime. Section \textbf{3} provides solutions of the field
equations by using three types of assumptions for particular values
of the parameters, given in the form of tables. In the last section,
we discuss and summarize the results.

\section{Field Equations in $f(R)$ Gravity for Cylindrically Symmetric Spacetime}

The modified form of the Einstein-Hilbert action in $f(R)$
gravity is given by \cite{34}
\begin{equation}\label{2.1}
S=\int d^{4}x\sqrt{-g}\left[f(R)+\kappa \mathcal{L_{\textit{m}}}\right],
\end{equation}
where $g$ is the trace of the metric tensor $g_{\mu\nu}$, $f(R)$ is
a generic function of the Ricci scalar $R,~\kappa$ is the coupling
constant in gravitational units and $\mathcal{L_{\textit{m}}}$ is
the standard matter Lagrangian. Variation of the action with respect
to the metric tensor leads to following  fourth order partial
differential equations
\begin{equation}\label{2.2}
F(R)R_{\mu\nu}-\frac{1}{2}f(R)g_{\mu\nu}-\nabla_{\mu}
\nabla_{\nu}F(R)+g_{\mu\nu}\Box F(R)=\kappa T_{\mu\nu},
\end{equation}
where $T_{\mu\nu}$ is the energy-momentum tensor of matter,
$F(R)\equiv df(R)/dR,~\Box \equiv \nabla^{\mu}\nabla_{\mu}$ with
$\nabla_{\mu}$ representing the covariant derivative. Taking trace
of the above equation, we obtain
\begin{equation}\label{2.3}
F(R) R - 2f(R)+ 3\Box F(R)= \kappa T.
\end{equation}
Here $R$ and $T$ are related differentially and not algebraically.
This indicates that the field equations of $f(R)$ gravity will admit
a larger variety of solutions as compared to GR.  Notice that
$R=0$ implies $T=0$ in GR whereas in $f(R)$ theory, this does not hold.
The Ricci scalar curvature function $f(R)$ can be expressed in terms
of its derivatives as
\begin{equation}\label{2.4}
f(R)=\frac{-{\kappa}T+F(R)R+3\Box F(R)}{2}.
\end{equation}
Substituting this value of $f(R)$ in Eq.(\ref{2.2}), we obtain
\begin{equation}\label{2.5}
R_{\alpha\beta}F(R)-\frac{1}{4}[{R F(R)-\Box
F(R)}]g_{\alpha\beta}-\nabla_{\alpha} \nabla_{\beta}F(R)= \kappa
[T_{\alpha \beta}-\frac{1}{4}(\rho-3p)g_{\alpha\beta}].
\end{equation}
The obtained equation is independent of $f(R)$ and is used to
calculate the independent field equations of a system.

The line element of static cylindrically symmetric spacetime is
given by \cite{S2}
\begin{equation}\label{3.1}
ds^2=A(r)dt^2-B(r)dr^2-r^2(d\theta^2+\alpha^2 dz^2),
\end{equation}
where $A,~B$ are the metric coefficients and $\alpha$ has dimensions of
$1/r$. The energy-momentum tensor for perfect fluid is
\begin{equation}\label{3.4}
T_{\mu\nu}=(\rho+p) u_\mu u_\nu- p g_{\mu\nu},
\end{equation}
here $u_\mu=\delta^0_\mu$ is the four-velocity in co-moving
coordinates, $\rho$ is the density and $p$ is the pressure of the
fluid. The corresponding Ricci scalar has the form
\begin{equation}\label{3.3}
R=\frac{1}{X}\left[{A''}+\frac{4A'}{r}-\frac{X'}{X}\left(\frac{A'}{2}+\frac{2A}{r}\right)
+\frac{2A}{r^2}\right],
\end{equation}
where $X={A}{B}$ and prime denotes derivative with respect to the
radial coordinate $r$. Using Eq.(\ref{2.5}), we get the following
two independent equations
\begin{eqnarray} \label{3.5}
-{2rF''}+r\frac{X'}{X}F'+2F\frac{X'}{X}-\frac{2r\kappa X}{A}(\rho+p)&=&0,\\\label{3.6}
{A''}+\left(\frac{F'}{F}-\frac{X'}{X}\right)\left({A'}-\frac{2A}{r}\right)
-\frac{2A}{r^2}-\frac{2\kappa X}{F}(\rho+p)&=&0.
\end{eqnarray}
The conservation equation,  $T^{\nu}_{\mu;\nu}=0$, leads to
\begin{equation}\label{3.7}
\frac{A'}{A}=-\frac{2p'}{\rho+p}.
\end{equation}
Thus we have a system of three non-linear differential equations
with five unknown functions, namely, $F(r),~\rho(r),~p(r),~A(r)$ and
$B(r)$.

\section{Solutions of the Field Equations}

This section is devoted to discuss the solutions of the field
equations. The Ricci scalar depends on $A(r)$ and $B(r)$, so it
would be much difficult to solve the field equations involving
$f(R)$. Consequently, analytical solutions of
Eqs.(\ref{3.5})-(\ref{3.7}) could not be found in closed form. Here,
we use Tolman's method \cite{35} which constructs a variety of
explicit solutions of the Einstein field equations. Tolman
introduced a relation for the metric components depending on radial
coordinates to solve the set of equations for perfect fluid. This
procedure yields pressure and density such that they may or may not
be physically acceptable. We explore cylindrically symmetric exact
solutions in terms of energy density, pressure and functions of $R$.
In order to solve the field equations, we use three types of
assumptions \cite{9,22} for $X$ and $F$ given below:
\begin{itemize}
\item $X=X_{0},~F=F_{0}r^{n}$,
\item $X=X_{0}r^m,~F=F_{0}$,
\item $X=X_{0}r^m,~F=F_{0}r^{n}$,
\end{itemize}
where $X_{0},~F_{0},~n$ and $m$ are arbitrary constants. For
any choice of $m$ and $n$, a number of interesting solutions
for $A,~\rho,~p$ and $R$ are obtained along with the
corresponding function $f(R)$.

\subsection{\textbf{Type I: $X=X_{0},~F=F_{0}r^{n}$}}

Inserting $X=X_{0},~F=F_{0}r^{n}$ in Eqs.(\ref{3.5}) and (\ref{3.6}) and
combining the resulting equations, we obtain
\begin{equation} \label{3.10}
{A''}+\frac{n}{r}A'+\frac{2n^2-4n-2}{r^2}A=0.
\end{equation}
Its general solution is
\begin{equation}\label{4.2}
A=A_{0}r^m,\quad m=\frac{1}{2}\left(1-n\pm
\sqrt{9+14n-7n^{2}}\right),
\end{equation}
where $A_{0}$ is a constant. Using this value in
Eqs.(\ref{3.5})-(\ref{3.7}), it follows that
\begin{equation}\label{4.3}
p=p_{0}+p_{1}r^{m+n-2},\quad p_{1}=\frac{F_{0}A_{0}mn(n-1)}{2\kappa
X_{0}(n+m-2)},
\end{equation}
where $p_{0}$ is another constant. Further
\begin{equation}\label{4.4}
\rho=-p_{0}+\rho_{1}r^{m+n-2},\quad
\rho_{1}=-\frac{F_{0}A_{0}n(n-1)(2n+3m-4)}{2\kappa X_{0}(n+m-2)}.
\end{equation}
The Ricci scalar takes the form
\begin{equation}\label{4.5}
R=R_{1}r^{m-2},\quad R_{1}=\frac{A_{0}}{X_{0}}(m^{2}+3m+2).
\end{equation}
We can write some exact solutions for particular values of $m$ and
$n$ given in \textbf{Table 1}. The last two columns have functional
form of $f(R)$ and speed of sound, respectively. We have already
assumed the isotropy of pressure. This type does not provide any
solution which satisfies the physically acceptable criteria
completely as the solutions I, III, V, VI and VII indicate squared
sound velocity as negative. The solutions II and IV do not have
monotonic decrease in energy density and pressure with increasing
radius.\\\\
\textbf{Table 1:} Type I solutions with
$X=X_{0},~F=F_{0}r^{n}$.
\begin{table}[bht]
\centering
\begin{small}
\begin{tabular}{|c|c|c|c|c|c|c|c|c|}
\hline\textbf{$No.$}&\textbf{$m$}&\textbf{$n$}
&\textbf{$A(r)$}&\textbf{${p}(r)$}&\textbf{$\rho(r)$}
&\textbf{$R(r)$}&\textbf{$f(R)$}&\textbf{${v_{s}}^{2}=\frac{dp}{d\rho}$}\\
\hline\textbf{$I$}&-3&1&\textbf{$A_{0}r$}
&\textbf{$p_{0}+\frac{A_{0}F_{0}r}{\kappa X_{0}}$}
&\textbf{$-p_{0}-\frac{3A_{0}F_{0}r}{\kappa X_{0}}$}
&\textbf{$\frac{6A_{0}}{X_{0}r}$}&\textbf{$f_{0}-\frac{5F_{0}R}{12}$}
&$-\frac{1}{3}$\\
\hline\textbf{$II$}&-2&0&\textbf{$A_{0}r^{2}$}
&$p_{0}$&$-p_{0}$& \textbf{$\frac{12A_{0}}{X_{0}}$}
&any regular&$<1$\\
\hline\textbf{$III$}&-2&1&\textbf{$\frac{A_{0}}{r^{2}}$}
&$p_{0}$ &$-p_{0}$&0&any regular&-1\\
\hline\textbf{$IV$}&0&1&\textbf{$A_{0}r$}
&$p_{0}$&$-p_{0}$&\textbf{$\frac{6A_{0}}{X_{0}r}$}
&any regular&$<1$\\
\hline\textbf{$V$}&0&2&\textbf{${A_{0}r^{2}}$}
&$p_{0}$&$-p_{0}$&\textbf{$\frac{12A_{0}}{X_{0}}$}
&any regular&-1\\
\hline\textbf{$VI$}&2.5&1&\textbf{$A_{0}r$}
&\textbf{$p_{0}+\frac{5A_{0}F_{0}r^{3/2}}{4\kappa X_{0}}$}
&\textbf{$-p_{0}+\frac{15A_{0}F_{0}r^{3/2}}{2\kappa X_{0}}$}
&\textbf{$\frac{6A_{0}}{X_{0}r^{3}}$}&\textbf{$f_{0}-\frac{F_{0}(3R^{-1/3}-5R)^{1/2}}{6}$}
&$-\frac{1}{4}$\\
\hline\textbf{$VII$}&2.5&2&\textbf{$\frac{A_{0}}{r}$}
&\textbf{$p_{0}-\frac{15A_{0}F_{0}}{4\kappa X_{0}\sqrt{r}}$}
&\textbf{$-p_{0}+\frac{15A_{0}F_{0}}{4\kappa X_{0}\sqrt{r}}$}
&0&any regular&$-\frac{1}{2}$\\
\hline
\end{tabular}
\end{small}
\end{table}

\subsection{\textbf{Type  II: \textbf{$X=X_{0}r^m,~F=F_{0}$}}}

Now replacing these values in Eqs.(\ref{3.5}) and (\ref{3.6}), it
follows that
\begin{equation}\label{4.6}
{A''}-\frac{m}{2r}A'-\frac{2+m}{r^2}A=0,
\end{equation}
which leads to the following solution
\begin{equation}\label{4.7}
A=A_{0}r^n,\quad n=\frac{1}{4}\left(2+m\pm
\sqrt{m^{2}+20m+36}\right).
\end{equation}
Using this solution, we can write pressure, density and Ricci scalar
as
\begin{eqnarray}\label{4.8}
p&=&p_{0}+p_{1}r^{n-m-2},\quad p_{1}=-\frac{F_{0}A_{0}mn}{2\kappa
X_{0}(n-m-2)}, \label{4.9}\\
\rho&=&-p_{0}+\rho_{1}r^{n-m-2},\quad\rho_{1}=\frac{F_{0}A_{0}m(3n-2m-4)}{2\kappa
X_{0}(n-m-2)}, \label{4.10}\\
R&=&R_{1}r^{n-m-2},\quad
R_{1}=\frac{A_{0}}{X_{0}}(n^{2}+3n-2m-nm+2).
\end{eqnarray}
For the same choices of $m$ and $n$ taken for type I solutions, we
construct a \textbf{Table 2} for this type of solutions given below.
Following the physical acceptability criteria, we see that solutions
II and V give negative squared sound velocity, while it becomes $1$
for solution IV. Solutions I and III do not show decrement in energy
density and pressure by increasing the value of radius. Only
solution VI satisfies complete criteria but the Ricci scalar turns
out to be radial dependent. Hence, we could not obtain any
acceptable solution for the constant curvature condition.\\\\
\textbf{Table 2:} Type II solutions with $X=X_{0}r^{m},~F=F_{0}$.
\begin{table}[bht]
\centering
\begin{small}
\begin{tabular}{|c|c|c|c|c|c|c|c|c|}
\hline\textbf{$No.$}&\textbf{$m$}&\textbf{$n$}& \textbf{$A(r)$}&
\textbf{${p}(r)$}&\textbf{$\rho(r)$}&\textbf{$R(r)$}&\textbf{$f(R)$}
&\textbf{${v_{s}}^{2}=\frac{dp}{d\rho}$}\\
\hline\textbf{$I$}&-2&0&\textbf{$A_{0}$}
&$p_{0}$&$-p_{0}$&\textbf{$\frac{6A_{0}}{X_{0}}$}
&\textbf{$f_{0}-\frac{F_{0}R}{2}$}&$<1$\\
\hline\textbf{$II$}&-2&1& \textbf{$A_{0}r$}
&\textbf{$p_{0}+\frac{A_{0}F_{0}r}{\kappa X_{0}}$}
&\textbf{$-p_{0}-\frac{3A_{0}F_{0}r}{\kappa X_{0}}$}
&\textbf{$\frac{12A_{0}r}{X_{0}}$}&\textbf{$f_{0}+\frac{5F_{0}R}{12}$}
&$-\frac{1}{3}$\\
\hline\textbf{$III$}&-3&1&\textbf{$A_{0}r$}
&\textbf{$p_{0}+\frac{3A_{0}F_{0}r^{2}}{4\kappa X_{0}}$}
&\textbf{$-p_{0}-\frac{15A_{0}F_{0}r^{2}}{4\kappa X_{0}}$}
&\textbf{$\frac{15A_{0}r^{2}}{X_{0}}$}&\textbf{$f_{0}-\frac{3F_{0}R}{10}$}
&$<1$\\
\hline\textbf{$IV$}&0&1&\textbf{$A_{0}r$}
&$p_{0}$&$-p_{0}$&\textbf{$\frac{6A_{0}}{X_{0}r}$}
&any regular&1\\
\hline\textbf{$V$}&0&2&\textbf{${A_{0}r^{2}}$}
&$p_{0}$&$-p_{0}$&\textbf{$\frac{12A_{0}}{X_{0}}$}
&any regular&-1\\
\hline\textbf{$VI$}&2.5&2&\textbf{$A_{0}r^{2}$}
&\textbf{$p_{0}+\frac{A_{0}F_{0}}{4\kappa X_{0}r^{5/2}}$}
&\textbf{$-p_{0}+\frac{6A_{0}F_{0}}{5\kappa X_{0}r^{5/2}}$}
&\textbf{$\frac{2A_{0}}{X_{0}r^{5/2}}$}&\textbf{$f_{0}-\frac{23F_{0}R}{20}$}
&$<1$\\
\hline
\end{tabular}
\end{small}
\end{table}

\subsection{\textbf{Type III: $X=X_{0}r^m,~F=F_{0}r^{n}$}}

This type of assumption deals with radial dependent expressions both
for $X$ and $F$, i.e.,
\begin{eqnarray}\nonumber
X=X_{0}r^m \quad F=F_{0}r^{n}
\end{eqnarray}
Replacing these values in Eqs.(\ref{3.5}) and (\ref{3.6}), it
follows that
\begin{equation}\label{4.1*}
{A''}-\frac{2n-m}{2r}A'-\frac{2n^{2}-4n-m-2-nm}{r^2}A=0,
\end{equation}
which has the following general solution
\begin{equation}\label{4.2*}
A=A_{1}r^{l_{1}}+A_{2}r^{l_{2}},
\end{equation}
where $A_{1},~A_{2}$ are constants and
$l_{1},~l_{2}$ are given as
\begin{equation}\nonumber
l_{1}=\frac{1}{4}\left(m-2n+ \sqrt{32+16m+m^{2}+64n+12mn-28n^{2}}\right),
\end{equation}
\begin{equation}\nonumber
l_{2}=\frac{1}{4}\left(m-2n- \sqrt{32+16m+m^{2}+64n+12mn-28n^{2}}\right).
\end{equation}
Inserting the above values in Eqs.(\ref{3.5})-(\ref{3.7}), the
corresponding $p,~\rho $ and $R$ are obtained as
\begin{equation}\label{4.3*}
p=p_{0}+p_{1}r^{l_{1}+n-m-2}+p_{2}r^{l_{2}+n-m-2}
\end{equation}
with
\begin{equation}\nonumber
p_{1}=\frac{F_{0}A_{1}l_{1}(2n^{2}-2n-2m-mn)}{4\kappa
X_{0}(l_{1}+n-m-2)},\quad
p_{2}=\frac{F_{0}A_{2}l_{2}(2n^{2}-2n-2m-mn)}{4\kappa
X_{0}(l_{2}+n-m-2)}.
\end{equation}
Also,
\begin{equation}\label{4.4*}
\rho=-p_{0}+\rho_{1}r^{l_{1}+n-m-2}+\rho_{2}r^{l_{2}+n-m-2},
\end{equation}
where
\begin{eqnarray*}
\rho_{1}&=&-\frac{F_{0}A_{1}(3l_{1}+2n-2m-4)(2n^{2}-2n-2m-mn)}{4\kappa
X_{0}(l_{1}+n-m-2)},\\
\rho_{2}&=&-\frac{F_{0}A_{2}(3l_{2}+2n-2m-4)(2n^{2}-2n-2m-mn)}{4\kappa
X_{0}(l_{2}+n-m-2)}.
\end{eqnarray*}
The corresponding Ricci scalar is
\begin{equation}\label{4.5}
R=R_{1}r^{l_{1}-m-2}+R_{2}r^{l_{2}-m-2},
\end{equation}
where
\begin{equation}\nonumber
R_{1}=\frac{A_{1}}{2X_{0}}(2l_{1}^{2}+(6-m)l_{1}+4m),\quad
R_{2}=\frac{A_{2}}{2X_{0}}(2l_{2}^{2}+(6-m)l_{2}+4m).
\end{equation}
We formulate solutions corresponding to $l_{1}$ and $l_{2}$ in two
different tables. Firstly, we calculate solutions for $l_{1}$ given
in \textbf{Table 3}.\\\\
\textbf{Table 3:} Type III solutions with
$X=X_{0}r^{m},~F=F_{0}r^{n}$ corresponding to $l_{1}$.
\begin{table}[bht]
\centering
\begin{small}
\begin{tabular}{|c|c|c|c|c|c|c|p{1in}|c|}
\hline\textbf{$No.$}&\textbf{$m$}
&\textbf{$n$}&\textbf{$A(r)$}&\textbf{${p}(r)$}
&\textbf{$\rho(r)$}&\textbf{$R(r)$}
&\textbf{$f(R)$}&\textbf{${v_{s}}^{2}=\frac{dp}{d\rho}$}\\
\hline\textbf{$I$}&-12&-1&\textbf{$\frac{A_{1}}{r}$}
&\textbf{$p_{0}-\frac{A_{1}F_{0}r^{8}}{2\kappa X_{0}}$}
&\textbf{$-p_{0}-\frac{15A_{1}F_{0}r^{8}}{2\kappa X_{0}}$}
&\textbf{$\frac{-32A_{1}r^{9}}{X_{0}}$}
&\textbf{$f_{0}-aF_{0}R^{8/9}$} where
\textbf{$a=\frac{1}{2^{5/9}}(\frac{A_{1}}{X_{0}})^{1/9}$}
&$-\frac{1}{8}$\\
\hline\textbf{$II$}&-10&-1&\textbf{$\frac{A_{1}}{r^{2}}$}
&\textbf{$p_{0}-\frac{7A_{1}F_{0}r^{5}}{5\kappa X_{0}}$}
&\textbf{$-p_{0}-\frac{28A_{1}F_{0}r^{5}}{5\kappa X_{0}}$}
&\textbf{$\frac{-32A_{1}r^{6}}{X_{0}}$}
&\textbf{$f_{0}-abF_{0}{R^{5/6}}$} where
\textbf{ $a=\frac{117}{5(2)^{25/6}}$},
\textbf{$b=(\frac{A_{1}}{X_{0}})^{1/6}$}&$-\frac{1}{4}$\\
\hline\textbf{$III$}&-2&0&\textbf{$A_{1}$}&$p_{0}$&$-p_{0}$
&\textbf{$\frac{-4A_{1}}{X_{0}}$}&any&$<1$\\
\hline\textbf{$IV$}&1&3&\textbf{$A_{1}$}&$p_{0}$&$-p_{0}$
&\textbf{$\frac{2A_{1}}{X_{0}r^{3}}$}&any regular&$<1$\\
\hline\textbf{$V$}&4&1&\textbf{${A_{1}r^{4}}$}
&\textbf{$p_{0}+\frac{12A_{1}F_{0}r^{-1}}{\kappa X_{0}}$}
&\textbf{$-p_{0}-\frac{6A_{1}F_{0}r^{-1}}{\kappa X_{0}}$}
&\textbf{$\frac{28A_{1}}{X_{0}r^{2}}$}
&\textbf{$f_{0}-\frac{17}{4\sqrt{7}}{aF_{0}}\sqrt{R}$} where
\textbf{$a=\sqrt{\frac{A_{1}}{X_{0}}}$}&2\\
\hline\textbf{$VI$}&6&2&\textbf{${A_{1}r^{5}}$}
&\textbf{$p_{0}+\frac{25A_{1}F_{0}r^{-1}}{\kappa X_{0}}$}
&\textbf{$-p_{0}-\frac{15A_{1}F_{0}r^{-1}}{\kappa X_{0}}$}
&\textbf{$\frac{37A_{1}}{X_{0}r^{3}}$}
&\textbf{$f_{0}+abF_{0}\sqrt[3]{R}$} where
\textbf{$a=\frac{63}{2\sqrt[3]{37}}$},
\textbf{$b=(\frac{A_{1}}{X_{0}})^{2/3}$}&$<1$\\
\hline\textbf{$VII$}&8&5&\textbf{$A_{1}r^{4}$}
&\textbf{$p_{0}+\frac{16A_{1}F_{0}r^{-1}}{\kappa X_{0}}$}
&\textbf{$-p_{0}-\frac{8A_{1}F_{0}r^{-1}}{\kappa X_{0}}$}
&\textbf{$\frac{26A_{1}}{X_{0}r^{6}}$}
&\textbf{$f_{0}+abF_{0}R^{1/6}$} where \textbf{
$a=\frac{51}{\sqrt[6]{2}\sqrt[6]{13}}$},
\textbf{$b=(\frac{A_{1}}{X_{0}})^{5/6}$}&2\\
\hline
\end{tabular}
\end{small}
\end{table}

Since we have already assumed isotropic pressure, so this condition
automatically holds for all the solutions. We see that solutions III
and IV have subluminal speed but energy density and pressure do not
decrease with the increase in radius. The solutions I and II show
decrease in density and pressure as radius increases but these give
negative squared sound velocity. For solutions V and VII, the energy
density and isotropic pressure decrease when the radius increases
and $v_{s}^{2}>1$. The solution VI fulfills the criteria of physical
acceptability properly. Hence this is the only acceptable
solution obtained by using this assumption.\\\\
\textbf{Table 4:} Type III solutions with
$X=X_{0}r^{m},~F=F_{0}r^{n}$ corresponding to $l_{2}$.\\
\begin{table}[bht]
\centering
\begin{small}
\begin{tabular}{|c|c|c|c|c|c|c|p{1in}|c|}
\hline\textbf{$No.$}&\textbf{$m$}&\textbf{$n$}&\textbf{$A(r)$}&
\textbf{${p}(r)$}&\textbf{$\rho(r)$}&
\textbf{$R(r)$}&\textbf{$f(R)$}&\textbf{${v_{s}}^{2}=\frac{dp}{d\rho}$}\\
\hline\textbf{$I$}&-2&0&\textbf{$\frac{A_{2}}{r}$}
&\textbf{$p_{0}+\frac{A_{2}F_{0}r^{-1}}{\kappa X_{0}}$}
&\textbf{$-p_{0}-\frac{3A_{2}F_{0}r^{-1}}{\kappa X_{0}}$}
&\textbf{$\frac{-7A_{2}}{X_{0}r}$}
&\textbf{$f_{0}-\frac{17}{14}R$}&$<1$\\
\hline\textbf{$II$}&4&1&\textbf{$\frac{A_{2}}{r^{3}}$}
&\textbf{$p_{0}-\frac{9A_{2}F_{0}r^{-8}}{8\kappa X_{0}}$}
&\textbf{$-p_{0}+\frac{57A_{2}F_{0}r^{-8}}{8\kappa X_{0}}$}
&\textbf{$\frac{14A_{2}}{X_{0}r^{9}}$}
&\textbf{$f_{0}-abF_{0}R^{(8/9)}$} where
\textbf{$a=\frac{91}{2^{26/9}\times7^{8/9}}$},
\textbf{$b=(\frac{A_{2}}{X_{0}})^{1/9}$}&$<1$\\
\hline\textbf{$III$}&4&3&\textbf{$\frac{A_{2}}{r^{4}}$}
&\textbf{$p_{0}-\frac{8A_{2}F_{0}r^{-7}}{7\kappa X_{0}}$}
&\textbf{$-p_{0}+\frac{36A_{2}F_{0}r^{-7}}{7\kappa X_{0}}$}
&\textbf{$\frac{-10A_{2}}{X_{0}r^{10}}$}
&\textbf{$f_{0}-abF_{0}R^{7/10}$} where
\textbf{$a=\frac{-205\times2^{3/10}}{7\times(5)^{7/10}}$},
\textbf{$b=(\frac{A_{2}}{X_{0}})^{3/10}$}&$<1$\\
\hline\textbf{$IV$}&6&2&\textbf{$\frac{A_{2}}{r^{4}}$}
&\textbf{$p_{0}-\frac{2A_{2}F_{0}r^{-10}}{\kappa X_{0}}$}
&\textbf{$-p_{0}+\frac{12A_{2}F_{0}r^{-10}}{\kappa X_{0}}$}
&\textbf{$\frac{28A_{2}}{X_{0}r^{12}}$}
&\textbf{$f_{0}-abF_{0}R^{5/6}$} where
\textbf{$a=\frac{3\times2^{7/3}}{7^{5/6}}$},
\textbf{$b=(\frac{A_{2}}{X_{0}})^{1/6}$}&$<1$\\
\hline\textbf{$V$}&6&5&\textbf{$\frac{A_{2}}{r^{4}}$}
&\textbf{$p_{0}-\frac{2A_{2}F_{0}r^{-7}}{7\kappa X_{0}}$}
&\textbf{$-p_{0}+\frac{9A_{2}F_{0}r^{-7}}{7\kappa X_{0}}$}
&\textbf{$\frac{28A_{2}}{X_{0}r^{12}}$}
&\textbf{$f_{0}-abF_{0}R^{7/12}$} where
\textbf{$a=\frac{165}{2^{1/6}\times7^{19/12}}$},
\textbf{$b=(\frac{A_{2}}{X_{0}})^{5/12}$}&$<1$\\
\hline\textbf{$VI$}&8&5&\textbf{$\frac{A_{2}}{r^{5}}$}
&\textbf{$p_{0}-\frac{2A_{2}F_{0}r^{-10}}{\kappa X_{0}}$}
&\textbf{$-p_{0}+\frac{10A_{2}F_{0}r^{-10}}{\kappa X_{0}}$}
&\textbf{$\frac{46A_{2}}{X_{0}r^{15}}$}
&\textbf{$f_{0}+abF_{0}\sqrt[3]{R^{2}}$} where
\textbf{$a=\frac{123}{{2}^{5/3}\times(23)^{2/3}}$},
\textbf{$b=(\frac{A_{2}}{X_{0}})^{1/3}$}&$<1$\\
\hline
\end{tabular}
\end{small}
\end{table}

The solutions corresponding to $l_{2}$ are shown in \textbf{Table
4}. In this table, the increment of radius monotonically decreases
pressure and density. Moreover, sound speed is subluminal. Thus we
can conclude that the most acceptable solution of a cylindrical star
in $f(R)$ gravity is generally described as:
\begin{eqnarray*}
A(r)&=&A_{2}r^{l},\\
p(r)&=&p_{0}+p_{2}r^{l+n-m-2},\\
\rho(r)&=&-p_{0}+\rho_{2}r^{l+n-m-2},\\
R(r)&=&R_{2}r^{l-m-2},\quad R_{2}=\frac{A_{2}}{2X_{0}}(2l^{2}+(6-m)l+4m),\\
f(R)&=&f_{0}+abF_{0}R^{1+k},\quad k=\frac{n}{l-m-2}.\\
a&=&\frac{1}{2}\left\{(2l^{2}+(6-m)l+4m)-\frac{3n}{2}(l-2m-4)-2n(n-1)\right.\nonumber\\
&-&\left.\frac{(10l+6n-6m-4)(2n^{2}-2n-2m-mn)}
{l+n-m-2}\right\}\{2l^{2}+(6-m)l+4m\}^{-(1+k)},\nonumber
\end{eqnarray*}
\begin{eqnarray*}
b&=&\left(\frac{A_{2}}{X_{0}}\right)^{-k}\nonumber,\quad
l=\frac{1}{4}\left(m-2n
-\sqrt{32+16m+m^{2}+64n+12mn-28n^{2}}\right).
\end{eqnarray*}

\section{Summary}

This paper is devoted to study the static cylindrically symmetric
solutions in metric $f(R)$ gravity for perfect fluid. In this
theory, the field equations are highly non-linear and complicated
which cannot be handled analytically without taking some
assumptions. For this purpose, we confine ourselves to only those
solutions which are closed and are found analytically without use of
any numerical data, perturbation or approximation method. The
corresponding form of $f(R)$ is reconstructed which does not depend
algebraically on $R(r)$ and $f(r)$. Physical acceptability criteria
is applied to check the physically acceptable solutions. We can
summarize the results as follows:
\begin{itemize}
\item  First type of assumption provides some solutions given in \textbf{Table 1} but none of
them satisfies physical acceptability measures. Thus no new solution
is obtained.
\item Second type is based on constant scalar curvature condition
which gives some new solutions in \textbf{Table 2}. It is noted that
some of them are physically acceptable but do not fulfill the
imposed conditions.
\item For the third family of solutions, we formulate two tables
corresponding to the parameters $l_{1}$ and $l_{2}$. We obtain one
solution in \textbf{Table 3} which satisfies the physically
acceptable criterion, whereas all the solutions in \textbf{Table 4}
fulfill the conditions for a physically acceptable solution. The
later gives the static cylindrically symmetric interior solutions in
the general form describing properties of a cylindrical star for
$f(R)$ gravity.
\end{itemize}

Since the exact solutions of $f(R)$ cosmological models are used to
explore dark energy and dust matter phases. Thus, these solutions
may provide a pointer towards the unknown nature of dark energy and
dark matter. These solutions might be useful at some stage to
overcome the theoretical difficulties in the context of cosmology
and astrophysics. The applicability of the solutions could be tested
by comparing with cosmology and some other constraints. It would be
interesting to investigate solutions for non-static spacetimes and
also with different types of fluid.

\end{document}